\def\msun{\ifmmode{\ {\rm M}_\odot}\else{$ {\rm M}_\odot$}\fi}
\def\msunyr{\ifmmode{\msun \ {\rm yr}^{-1}}\else{$\msun \ {\rm yr}^{-1}$}\fi}
\newcommand{\reff}{r_{\rm eff}}
\newcommand{\zmin}{z_{\rm min}}
\newcommand{\zmax}{z_{\rm max}}

\documentclass{emulateapj}

\usepackage{apjfonts}
\usepackage{graphicx}

\slugcomment{Submitted for publication in the Astrophysical Journal}

\shorttitle{Supercavities in Hydra A}
\shortauthors{Wise et al.}

\begin{document}

\title{X-ray Supercavities in the Hydra A Cluster and the Outburst History
of the Central Galaxy's Active Nucleus}

\author{M. W. Wise\altaffilmark{1,2},
         B. R. McNamara\altaffilmark{3,4,5},
         P. E. J. Nulsen\altaffilmark{5,6},
         J. C. Houck\altaffilmark{2},
         and L. P. David\altaffilmark{5}}



\altaffiltext{1}{Astronomical Institute ``Anton Pannekoek'', 
                 University of Amsterdam,
                 Kruislaan 403, 1098 SJ Amsterdam, The Netherlands}
\altaffiltext{2}{Massachusetts Institute of Technology, Kavli Institute,
                  Cambridge, MA 02139--4307}
\altaffiltext{3}{Department of Physics \& Astronomy, University of  
                 Waterloo, Ontario, Canada}
\altaffiltext{4}{Astrophysical Institute and Dept. of Physics \&  
                 Astronomy, Ohio University, Athens, OH, 45701}
\altaffiltext{5}{Harvard-Smithsonian Center for Astrophysics,
                 60 Garden Street, Cambridge, MA  02138}
\altaffiltext{6}{On leave from the University of Wollongong, Australia}


\begin{abstract}
A 227 ksec {\it Chandra} Observatory X-ray image of the hot plasma
in the Hydra A cluster has revealed an extensive cavity system.
The system was created by a continuous outflow or a series of bursts
from the nucleus of the central galaxy over the past $200-500$ Myr.
The cavities have displaced 10\% of the plasma within a 300 kpc radius
of the central galaxy, creating a swiss-cheese-like topology in the
hot gas. The surface brightness decrements are consistent with
empty cavities oriented within 40 degrees of the plane of the sky.
The outflow has deposited upward of $10^{61}$ erg
into the cluster gas, most of which was propelled
beyond the inner $\sim 100$ kpc cooling region.
The supermassive black hole has
accreted at a rate of approximately $0.1-0.25 \msunyr$ over this
time frame, which is a small fraction of the Eddington rate of a
$\sim 10^9\msun$ black hole, but
is dramatically larger than the Bondi rate. 
Given the previous evidence for a circumnuclear disk of cold gas in
Hydra A, these results are consistent with the AGN 
being powered primarily by infalling cold gas.   
The cavity system is shadowed perfectly by 330 MHz radio emission.
Such low frequency synchrotron emission may be an excellent
proxy for X-ray cavities and thus the total energy liberated by the
supermassive black hole. 
\end{abstract}



%
%
\section{Introduction}
\label{sec:intro}

The realization that powerful outbursts from central active galactic
nuclei (AGN) can release upward of $10^{61}$ erg into the intracluster
medium (ICM) points to a common solution to several heating problems
associated with clusters and groups of galaxies \citep{mcn05,nul05a,nul05}.  
AGN feedback has become the most promising mechanism for regulating
the cooling of the hot gas in galaxies and clusters, and thus limiting
the growth of elliptical galaxies and their attendant supermassive
black holes \citep{vd05,bbf03}. Furthermore, energy released by AGN
could be a major source of excess entropy  (``preheating'') in the hot
gas in groups and clusters \citep[][]{wfn00, voit04}, and AGN
outbursts may thread clusters with large-scale magnetic fields. 

Cavities and shock fronts provide a reliable lower limit to
the energy output of AGN independent of the magnetic field and particle
content of the jet \citep{birz04, dft05, deyoung06}.
The existence of well-defined cavities in the hot ICM around radio lobes
implies that the magnetic field and relativistic particles are largely
confined to the volume of the cavity such that most of the energy
output of the AGN is included (radiation losses are usually negligible
in these systems).  This energy provides a strict lower limit to
the gravitational binding energy released by accretion
onto the embedded supermassive black hole, and hence the mean growth
rate of the hole over the life of the outburst \citep 
{mcn05,nul05a,rmn06}.

This geometric approach is limited at present by
the {\it Chandra} X-ray Observatory's
ability to resolve cavity boundaries required to measure their
sizes (volumes) and the surrounding pressures.  This provides a largely
parameter free estimate of the work done by radio jets as they
inflate the cavities
against the surrounding gas pressure ($pV$).  The method
assumes the cavities are close to pressure balance
with the surrounding gas, which is justified by the absence of
  bounding strong shocks, their high detection rate and advanced
ages \citep{birz04, dft05}.
The total free energy (enthalpy) per cavity
is expected to lie in the range $2.5pV$ -- $4pV$ if the pressure is
dominated by gas, or $2pV$ if it is dominated by magnetic field.
Thus, the free energy depends
on the equation of state of the (radio) plasma filling the cavities.
The principal sources of uncertainty in the enthalpy estimates are  
the volume
measurements (geometry) and the as yet unknown makeup
(equation of state) of the plasma inside the cavities,
which combined amount to uncertainties of factors of a few.

%
%
\begin{figure*}[t]
\centering
\includegraphics[width=6.75in]{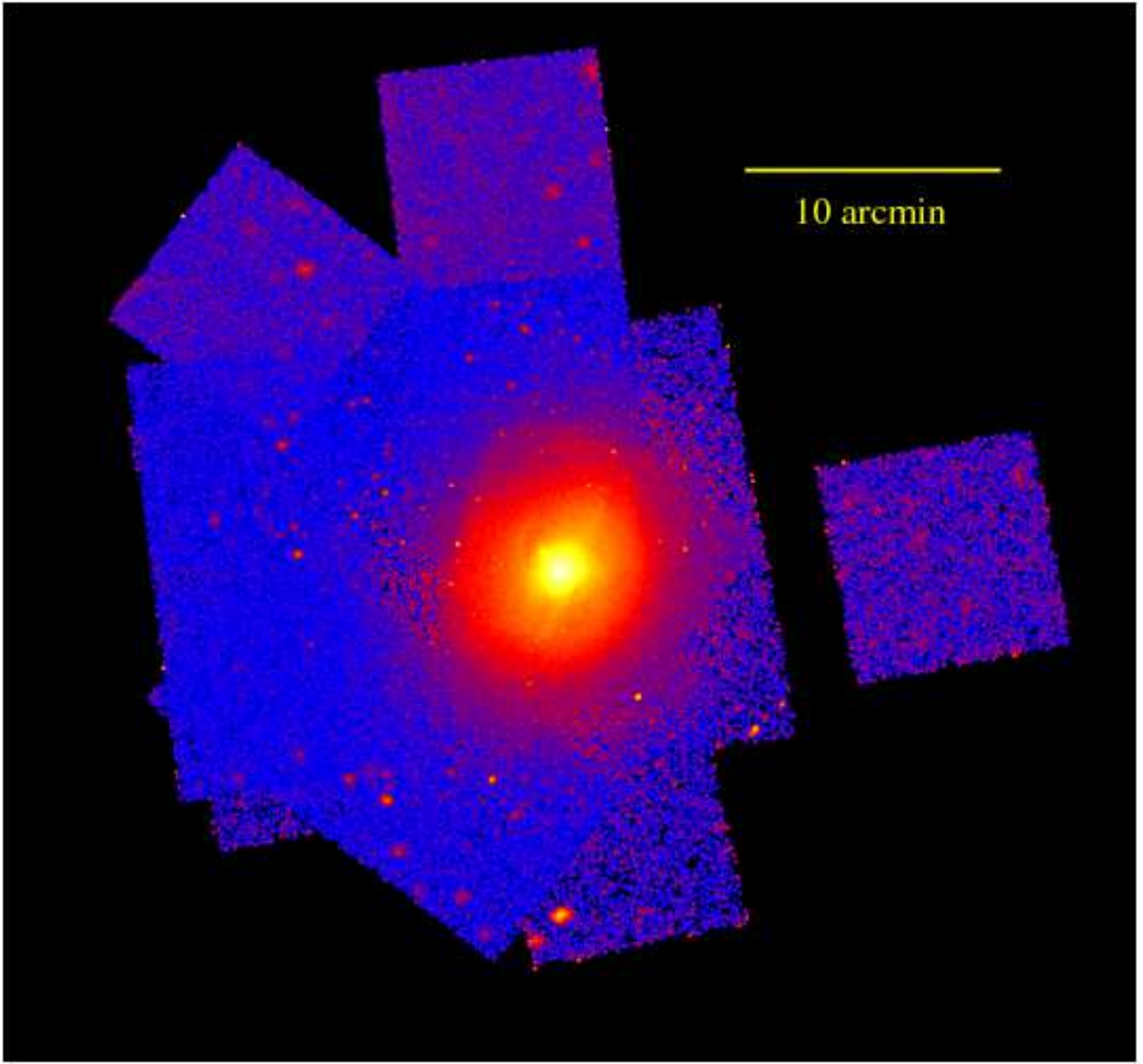}
\caption{ {\it Chandra} 0.5-7.0 keV mosaic of the Hydra A field.
The image has been background-subtracted and exposure-corrected before
smoothing with a $1.5 \arcsec$ Gaussian. The horizontal bar shows the
spatial scale in the image. A logarithmic color table has been used to
highlight low surface brightness features.}
\label{fig:mosaic}
\end{figure*}

The total energy pumped into the ICM by AGN is partitioned
into radio emission (which is
generally negligible), cavity energy
($3pV$ for relativistic gas), and ``shock energy.''
Here ``shock energy'' refers to the total work done by the expanding
cavities on the surrounding gas, which appears as excess thermal,
kinetic, and potential energy in the ICM.  Its effect is most evident
where the expanding cavities drive a detectable shock.
Measurements of these quantities have
shown that the internal cavity pressures
generally exceed the radio equipartition pressures by more than an
order of magnitude and that jet (mechanical) power often exceeds
synchrotron power by several orders of magnitude.
The unavoidable conclusion then is that radio jets possess
much larger mechanical powers than expected,
with potentially dramatic consequences for quenching cooling flows
and cluster heating \citep{pgd00}.

Whether AGN heating of the ICM occurs violently through shocks or
outflows, or gently through bubbling, vorticity, etc., or both is
poorly understood.  Most known cavity systems are associated
with Fanaroff-Riley class I radio sources, for which jet ram pressure
is probably negligible.  In this instance the pressure within the
X-ray cavities is expected to be uniform \citep{hrb98}, and thus the 
cavity pressure is well-defined by the external pressure. The ratio of
cavity enthalpy to shock energy is governed chiefly by the history of
power input to the cavity. Therefore, the nature and history of energy 
injection can be evaluated using the cavity properties and the
surrounding shock front.  Here we examine this and other issues, using
an analysis of the enormous system of X-ray cavities embedded in a
``cocoon'' shock in the Hydra A cluster \citep{mcn00,dnm01,nul02,nul05}, 
and we discuss its consequences for cluster heating by the central radio 
source, and for accretion onto the central supermassive black hole.

In the succeeding analysis, we have adopted a redshift for Hydra A
of $z = 0.0538$ and a flat $\Lambda$CDM cosmology with
$\rm{H}_0$=70\hspace{.05in} $\rm{km\hspace{.05in}s^{-1}\hspace{.05in} 
Mpc^{-1}}$,
$\Omega_M$=0.3, and $\Omega_\Lambda$=0.7.
These assumptions yield a luminosity distance of $D = 240$ Mpc,
an angular diameter distance of 216 Mpc,
and a linear scale of 1.05 kpc per arcsec.

%
%
\begin{figure*}[ht]
\centering
\includegraphics[width=6.6in]{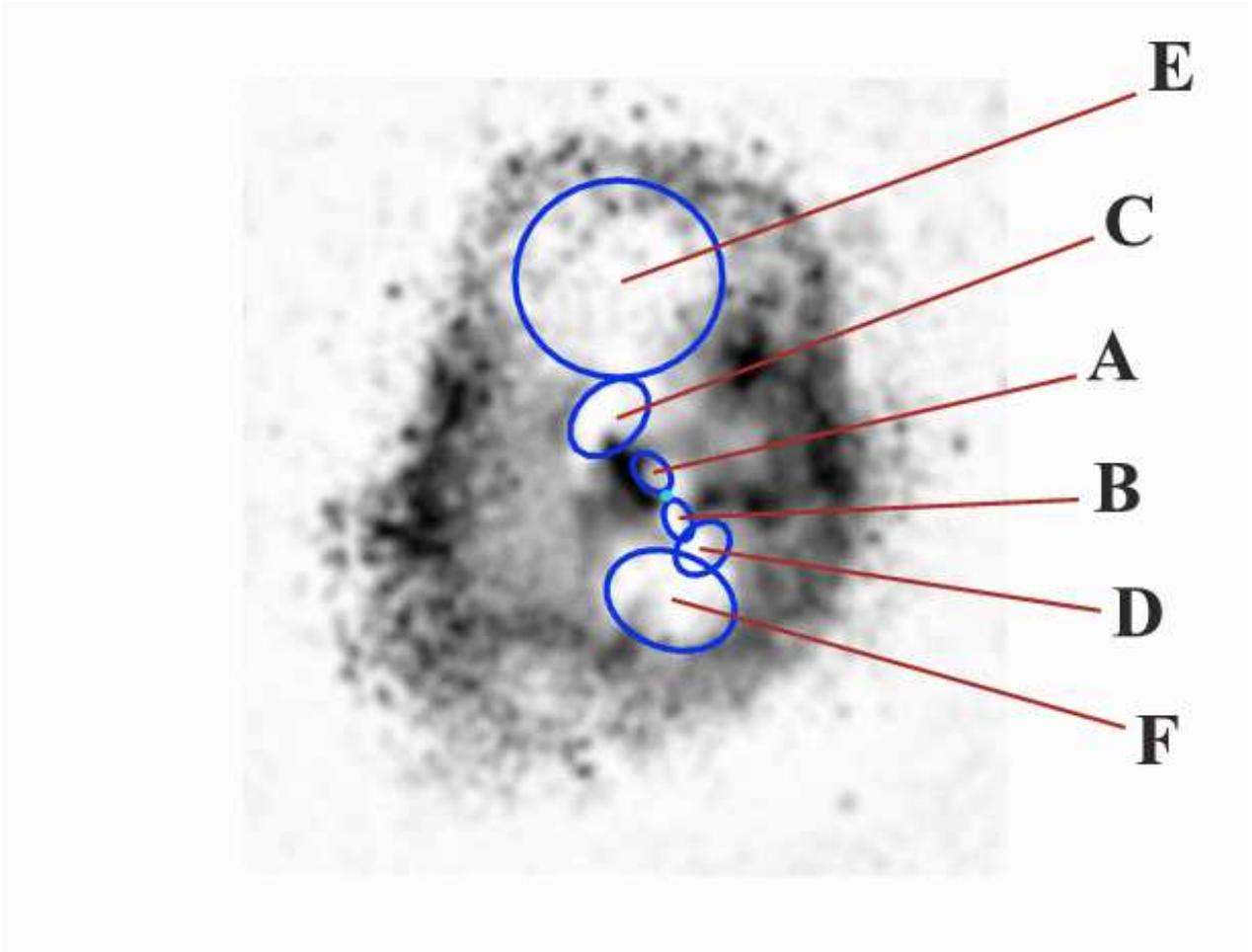}
\caption{High-contrast residual map after subtraction of a smooth,
elliptical beta model fit to the cluster surface brightness profile.
The positions of the three pairs of cavities discussed in the text
are indicated.The position of the central radio source is indicated
by the light blue dot.}
\label{fig:cavities}
\end{figure*}

\section{Observations and Data Reduction}
\label{sec:data}

Hydra A has been observed four times by {\it Chandra} with the ACIS
detector in imaging mode for a total exposure of 240 ksec.
Three of the observations were made with ACIS-S detector at
the aimpoint (ObsIDs 576, 4969, 4970) and one (ObsID 575)
using the ACIS-I detector.
The two shorter exposures were taken early in the {\it Chandra} mission
when the ACIS detector was operated at the higher focal plane
temperature of $-110^\circ$ C. Each dataset was individually
reprocessed using CIAO 3.2 and calibration files
available in CALDB 3.0. Standard screening was applied to the
event files to remove bad grades and pixels. Two of the datasets
(ObsIDs 575 and 4969) exhibited strong background flares and required
additional filtering. The final, combined exposure time for all four
datasets was 227 ksec.

After cleaning, the event files for ObsID's 575, 576, and 4969 were
reprojected to a common tangent point on the sky with ObsID 4970.
A number of bright point sources in Hydra A were used
to test for potential astrometry errors in the reprojected sky
coordinates between fields.  Individual background event files were
created for each ObsID from the standard ACIS blank-sky event files
following the procedure described in \cite{vikhlinin05} and 
reprojected to the common tangent for ObsID 4970.
Using the reprojected event files, source and background counts 
images were created for each CCD in each observation using standard
CIAO tools. Matching exposure maps were created for each chip
using the CIAO tool {\tt expmap}. The resulting images for individual
CCDs were then added to form mosaics combining all four datasets.
The resulting background-subtracted, exposure-corrected mosaic is
shown in Figure~\ref{fig:mosaic} centered on the radio source 3C 218.
All detected point sources were excised prior to extracting
spectra and surface brightness profiles. 

Due to the higher quality calibration at $-120^\circ$ C, spectral
analysis was restricted to the two longer, more recent exposures
(4969 and 4979).
Collectively, ObsIDs 575 and 576 comprise 17\% of the total exposure
and neglecting them has little or no effect on the final spectral fits.
Count weighted detector response (RMFs) and effective area (ARFs) files
were created for extraction regions using the CIAO tools
{\tt mkacisrmf} and {\tt mkwarf}, including the temporal, spectral,
and spatial dependences of the ACIS filter contaminant.

%
%
\begin{deluxetable*}{lrrrrrrrrc}[hb]
\tablecolumns{10}
\tablewidth{0pt}
\tablecaption{Cavity Properties in Hydra A}
\tablehead{
\colhead{~} &
\colhead{$a^{a}$} &
\colhead{$b^{b}$} &
\colhead{$R^{c}$} &
\colhead{$pV$} &
\colhead{$t_{c_{s}}^{d}$} &
\colhead{$t_{buoy}^{e}$} &
\colhead{$t_{r}^{f}$} &
\colhead{$\langle t \rangle^{g}$} &
\colhead{$P_{\rm cav}$$^{h}$} \\
\colhead{Cavity} &
\colhead{(kpc)} &
\colhead{(kpc)} &
\colhead{(kpc)} &
\colhead{($10^{58}$ ergs)} &
\colhead{($10^7$ yr)} &
\colhead{($10^7$ yr)} &
\colhead{($10^7$ yr)} &
\colhead{($10^7$ yr)} &
\colhead{($10^{44}$ ergs s$^{-1}$)} \\
}
\startdata
A &   20.5 &   12.4 &   24.9 &    8.4 &    2.7 &    4.3 &    8.6 &     
5.2 &    2.1 \\
B &   21.0 &   12.3 &   25.6 &    8.2 &    2.7 &    4.4 &    8.7 &     
5.3 &    2.0 \\
C &   47.2 &   31.5 &  100.8 &   29.3 &   10.2 &   23.0 &   26.8 &    
20.0 &    1.9 \\
D &   29.0 &   20.9 &   59.3 &   14.8 &    6.1 &   13.3 &   16.4 &    
11.9 &    1.6 \\
E &  105.0 &   99.7 &  225.6 &  247.9 &   22.3 &   51.7 &   65.3 &    
46.4 &    6.8 \\
F &   67.7 &   50.1 &  104.3 &  102.3 &   10.5 &   20.2 &   33.5 &    
21.4 &    6.1 \\
\enddata
\tablenotetext{a}{Projected semimajor axis of the cavity.}
\tablenotetext{b}{Projected semiminor axis of the cavity.}
\tablenotetext{c}{Projected distance from the cavity center to the radio core.}
\tablenotetext{d}{Estimated cavity age based on the sound speed, $c_s$, in the gas.}
\tablenotetext{e}{Estimated cavity age assuming the cavity rises buoyantly.}
\tablenotetext{f}{Estimated cavity age based on the refill timescale.}
\tablenotetext{g}{Mean cavity age for all three estimates.}
\tablenotetext{h}{Cavity power is calculated assuming $4 pV$ of energy
                  per cavity and the mean timescale for the age of the
                  cavity $\langle t \rangle$.} 
\label{tab:cavities}
\end{deluxetable*}

\section{Cluster-scale Radio Cavities \& Filling Factor}
\label{sec:bubbles}

In order to highlight faint cavity structures,
an elliptical beta model was fit to the azimuthally averaged surface
brightness and subtracted. The residual image shown in
Figure~\ref{fig:cavities} reveals at least
three pairs of X-ray cavities emanating from the central cD galaxy
distributed with a ``swiss cheese'' topology throughout the inner  
region of the cluster.

The inner pair of cavities (labeled A and B in Figure~\ref{fig:cavities})
are well-known set described by \citet{mcn00, dnm01, nul02}.
These cavities are each $40$--$50$ kpc in diameter and centered
roughly 25 kpc north-south of the central AGN. We also note that the
enhanced X-ray emission, relative to the underlying smooth model,
immediately to the west of cavity A corresponds to the region of
cooler gas seen previously in \cite{nul02}.

A second set of ``ghost'' cavities (labeled C and D) can be seen beyond
the inner pair. The northern-most corresponds to the cavity ``C''
feature discussed in \citet{nul05}. This asymmetrical pair of cavities
has diameters of $50$ kpc and $80$ kpc, projected 100 kpc and 60 kpc
distant from the north and south cavities, respectively.
Finally, two very large cavities (E and F) are projected at radii of
225 kpc and 100 kpc to the north and south, respectively. These
cavities have diameters of $120$--$200$ kpc and fall just inside a
sharp edge in the surface brightness which can also be seen in
Figure~\ref{fig:cavities}. This edge was interpreted by \citet{nul05}
as a shock front associated with a 140 Myr old outburst. The estimated
Mach number for this shock is roughly 1.2, although the data are not
of sufficient quality to detect the 20\% jump in temperature expected
for such a shock.

This enormous cavity system fills at least $\sim$10\% of the cluster's
volume within a 300 kpc radius (see below).
Table~\ref{tab:cavities} describes the cavities in the system.
All appear to be connected, raising the possibilities that they are
merging bubbles generated by a series of outbursts, or several
attached cavities that have been continuously powered by the AGN over
the past several hundred million years.

%
%
\begin{deluxetable*}{lrrrrrrrc}[ht]
\tablecolumns{9}
\tablewidth{0pt}
\tablecaption{Cavity Decrements in Hydra A}
\tablehead{
\colhead{~} &
\colhead{$\reff$$^{a}$} &
\colhead{$R$} &
\colhead{$y$$^{b}$} &
\colhead{$\delta y^{c}$} &
\colhead{$z$$^{d}$} &
\colhead{$\zmin$} &
\colhead{$\zmax$} &
\colhead{atan($z/R$)} \\
\colhead{Cavity} &
\colhead{(kpc)} &
\colhead{(kpc)} &
\colhead{~} &
\colhead{~} &
\colhead{(kpc)} &
\colhead{(kpc)} &
\colhead{(kpc)} &
\colhead{(deg)} \\
}
\startdata
A &  15.97 &  24.9 &   0.90 &   0.06 &   45 &  35 &  65 &  61  \\
B &  16.06 &  25.5 &   0.67 &   0.04 &   19 &  15 &  22 &  37  \\
C &  38.58 & 100.8 &   0.71 &   0.08 &   26 & $-$ &  70 &  14  \\
D &  24.62 &  59.3 &   0.59 &   0.05 &  $-$ & $-$ & $-$ & $-$  \\
E & 102.34 & 225.6 &   0.62 &   0.15 &  101 & $-$ & 193 &  24  \\
F &  58.26 & 104.3 &   0.71 &   0.09 &   80 &  43 & 118 &  37  \\
\enddata
\tablenotetext{a}{For ellipsoidal cavities, the effective radius is
   taken to be $\reff = (a b)^{1/2}$, with values taken from
   Table~\ref{tab:cavities}.}
\tablenotetext{b}{The cavity decrement as discussed in the text,
   expressed as the ratio of the surface brightness at the center of
   the cavity relative to the underlying smooth surface brightness 
   model, $y = S_{cav}/S_{smo}$. Values less than one show decrements.}
\tablenotetext{c}{Estimate of the $1\sigma$ error in the cavity
   decrement based on the width of the distribution of $y$ values as
   measured in a 5 arcsec radius, circular aperture centered on the
   cavity center.}
\tablenotetext{d}{Distance from the plane of the sky containing the
   AGN along the line-of-sight to the cluster.}
\label{tab:decrements}
\end{deluxetable*}

\section{Cavity Geometry \& Energetics}
\label{sec:energetics}

The deprojected gas densities and temperatures surrounding the
cavities were determined using the techniques described in
\citet{wise04} and are consistent with those of \cite{dnm01}.
Values for the surrounding pressures were combined with the
volume estimates to calculate the $1pV$ energies of each cavity
listed in Table~\ref{tab:cavities}.
Summing the $pV$ values for all six cavities from
Table~\ref{tab:cavities}, the total work done by the jet is 
$4.1\times 10^{60}$ ergs, corresponding to a total free energy
or enthalpy ($2pV$ -- 4$pV$) of 0.8 -- $1.6\times10^{61}$ ergs.
This range is comparable to the energetic output of other supercavity
systems MS0735.6+7421: $8\times10^{61}$ erg \citep{mcn05}, and
Hercules A: $3\times10^{61}$ erg \citep{nul05a}. It is also
comparable to the $\sim1\times10^{61}$ erg \citet{nul05} found 
for the energy associated with the large-scale shock in Hydra A.

Inspection of the cavity energies in Table~\ref{tab:cavities} shows
that the outer, larger cavities contain more than an order of  
magnitude more energy than the original A and B cavity systems.
Most of this energy is being deposited into the ICM outside the 100 kpc
cooling radius in Hydra A \citep{dnm01,birz04}. Less than 10\% of the
total free energy associated with the cavity system resides
inside the cluster's cooling radius.

A number of factors contribute to the uncertainty in these energy
estimates including geometric uncertainties in the size and shape
of the cavities, the ``filling factor'' or fraction of the volume 
from which X-ray emitting gas is excluded, and the locations
of the cavities along the line-of-sight through the cluster.
Since these factors also affect the observed X-ray surface brightness  
profile over the cavities, these profiles can be used to constrain
these uncertainties and to test our assumptions about the cavities.
For example, the presence of an empty cavity within the cluster
emitting volume will produce a decrement in the X-ray surface
brightness that depends on the size of the cavity and its depth 
in the cluster. By measuring these cavity decrements, we can therefore
set limits on the placement of the cavity along our line-of-sight
through the cluster.

Table~\ref{tab:decrements} gives the decrement for each cavity, $y$,
defined to be the ratio of the surface brightness measured at the
center of each cavity to the surface brightness at the same radius in
the ``undisturbed'' ICM. The undisturbed ISM is modeled using the
the double beta model fit to the azimuthally averaged surface
brightness profile (excluding the cavities) of \citet{dnm01}.  
This model is described by values of $\beta_1 = 0.686$, $\beta_2 = 0.907$, 
core radii of 27.7 kpc and 235.6 kpc, respectively, and a normalization 
for the more extended, second beta model component relative to the
first of 0.0418. To estimate the error in the decrement, $\delta y$, 
we determined the distribution of $y$ values in circular apertures
5 arcsec in radius, centered on the cavity centers. The decrement
errors, $\delta y$, quoted in Table~\ref{tab:decrements} represent
the $1\sigma$  widths of these distributions.

If we assume that the width of the cavity along the line of sight
is $2 \reff = 2 (ab)^{1/2}$ and the underlying surface brightness
profile is given by the double beta model fit described above, 
we can determine the distance, $z$, of the cavity along the line-of-sight.
This distance is measured from a plane perpendicular to the 
line-of-sight which passes through the central AGN.
Table~\ref{tab:decrements} lists the values of $z$ for each cavity
required to reproduce the observed surface brightness decrement $y$.
One sigma ranges for $z$, $\zmin$ -- $\zmax$ are also given.  
Missing entries fall outside the range of surface brightness allowed
by this model.  These values are insensitive to details of the surface
brightness model.  Using the single beta model of \citet{daf90} for
the outer cavities, or a single beta model for the {\it Chandra} data
at small radii for the inner cavities gives very similar results.

With the possible exception of cavity A, the results in
Table~\ref{tab:decrements} show that the the cavities must be almost
devoid of X-ray emitting gas at the ambient temperature and density
and lie close to the same plane as the AGN. If their filling factors
are less than unity, then they would need to lie even closer to the
plane of the AGN.  Thus, the surface brightness deficit for most of
the cavities is consistent with our key assumptions. 
Only the deficit of 0.1 for cavity A is marginally smaller than
expected. Its elongation along the northern radio ``jet'' suggests
that cavity A may be prolate; however, replacing the effective radius
with the semiminor axis in the model only shifts it a little closer to
the plane of the AGN, from 45 kpc  to 39 kpc.  
At a distance of 45 kpc, it makes an angle of $\sim60^\circ$ to the plane
of the sky, well within reasonable bounds.  However, the alignment of
the inner radio lobes (cavities A and B) in the plane of the sky
suggests that they should make the same angle to the plane of the sky.
Since the other cavities all make smaller angles to the plane of the
sky, cavity A appears to be the anomaly.  Its low decrement may be due
to dense, cool gas lying over the cavity.  In any case, cavity A
contributes little to the total energy.  

\begin{figure*}[t]
\centering
\includegraphics[width=6.5in]{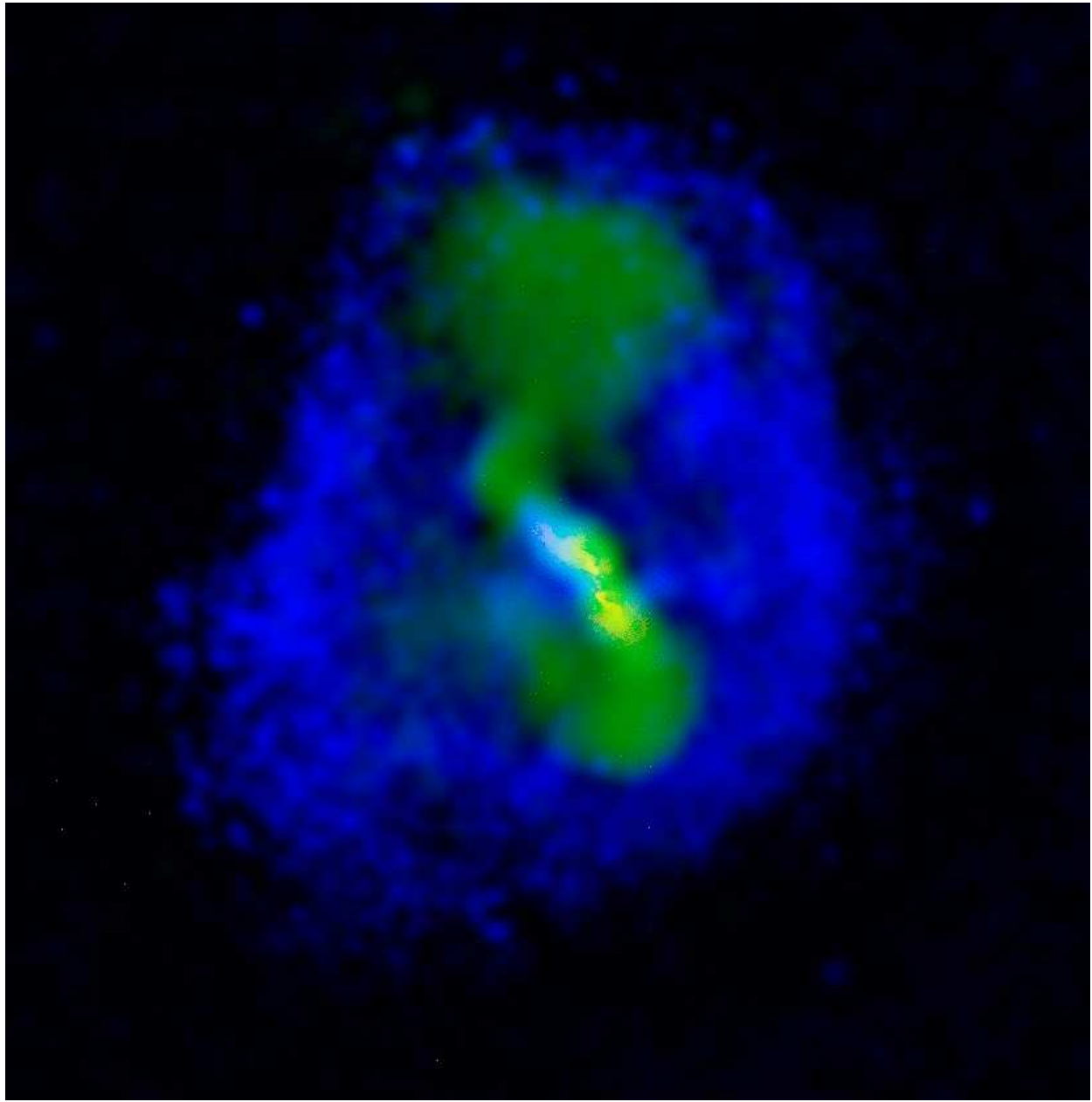}
\caption{Composite color image of Hydra A which illustrates the
close connection between the observed, large X-ray cavity system
(shown in blue) and the low frequency, 330 Mhz radio emission
(shown in green). The X-ray emission corresponds to the residual image
shown in Figure~\ref{fig:cavities}. The 330 Mhz radio data is from 
 \cite{lane04}. The familiar 1.4 GHz VLA image of Hydra A is also
shown in the core in yellow.}
\label{fig:radio}
\end{figure*}

By contrast, the deficit for cavity D is too large to be explained by
our model.  Cavity D lies where the southern radio lobe appears to
fold back on itself \citep[also Figure~\ref{fig:radio}]{lane04}, so that
our line-of-sight may be nearly tangent to the lobe axis there.  In
that case, it would be reasonable to expect its central depth to
exceed $2\reff$.  If true, the $pV$ value quoted for cavity D in
Table~\ref{tab:cavities} may be an underestimate.

%
%

\section{Cavity Ages and Powers}
\label{sec:ages}

Analogous to the well-known 1.4 Ghz radio emission filling the
innermost cavities, low frequency radio observations of Hydra A 
show that these large-scale cavities also contain relativistic plasma
strongly suggesting a connection to the AGN located in the nucleus of
the central cD galaxy. This point is illustrated in Figure~\ref{fig:radio} 
which overlays the 330 Mhz VLA image of \cite{lane04} with the
residual X-ray image shown in Figure~\ref{fig:cavities}. 
The existence of multiple cavities implies that the central AGN is
periodically rejuvenated producing a series of outbursts. 

We can gauge the timescale over which this activity has occurred
by determining the ages of these cavities. Following the discussion
in \cite{birz04}, we have estimated the age of each cavity in three
ways. First, the age of the cavity was estimated as the time required
for it to reach its projected location assuming it traveled at the
sound speed. Alternatively, the age was estimated as the time for the
cavity to rise buoyantly to its present location. Finally, the age was
taken to be the refill timescale \citep{mcn00,nul02}. Each of these
timescales is tabulated in Table~\ref{tab:cavities} along with the
mean cavity age, $\langle t \rangle$, for the three methods. 
Cavity ages based on the sound speed typically yield the smallest
timescales while refill timescales tend to be the
largest \cite[cf][]{birz04}. We find mean ages of $\sim 50$ Myr for
cavities A and B, consistent with \citep{birz04} and values ranging
between  $\sim100$ -- 200 Myr for cavities C and D.  Cavity E could be
up to almost 500 Myr old. 

Combining the values of $pV$ given in Table~\ref{tab:cavities} with
these age estimates, we can calculate the mean instantaneous power 
of each cavity as $P_{\rm cav} = 4pV/\langle t \rangle$. Here, we have 
explicitly assumed the cavity is filled predominantly with
relativistic plasma, making its enthalpy $4pV$. The resulting values
for $P_{\rm cav}$ are presented in Table~\ref{tab:cavities}.  
Equating $P_{\rm cav}$ with the output of the AGN, the data for the
four inner cavities imply that a pair of cavities is created in Hydra A 
every $\sim 50$ -- 100 Myr for a mean AGN power of $2 \times 10^{44}$ ergs s$^{-1}$.
On the face of it, the larger, outermost cavities would seem to be
associated with an earlier, more energetic state of activity which
occurred 200 -- 500 Myr ago with an associated AGN power output
of $\sim 6$ -- $7 \times 10^{44}$ ergs s$^{-1}$.

\section{Shock Front Age and Power}

By fitting simple shock models to the surface brightness discontinuity
in Hydra A observed at $\sim 210$ kpc, \cite{nul05} estimated an age
for the shock of 140 Myr, smaller than the shortest age of 220 Myr for
cavity E found here. At its closest point to the AGN, the projected
distance from the shock front to the AGN should be very nearly equal
to the actual distance, so that the age estimate based on the shock is
likely to be the most accurate.  Given their relative locations, it is
not surprising that an age based on the closest points of the
(supersonic) shock front is smaller than any estimate in
Table~\ref{tab:cavities} for the age of cavity E. This result is a
reminder that a momentum dominated jet can excavate a remote cavity in
considerably less time than it would take for the cavity to rise
buoyantly from the AGN.  It also illustrates that cavity based ages
may be too high, and the total power may be underestimated. 

The preceding interpretation presupposes that the cavities do not
share energy, whereas the radio and X-ray  maps show that they might
be interlinked (Figure~\ref{fig:radio}).  Furthermore, if the cavities
are disconnected, then the jet cannot supply power to the outermost
cavities that are chiefly responsible for driving the shock.  
We should then expect the shock to separate rapidly from those cavities.
For example, cavity E needs to be overpressured by a factor close to 2
in order to drive the Mach 1.34 shock to the north \citep{nul05}.  
If it is dominated by relativistic gas, then it can only expand
adiabatically by about 19 kpc from its current radius, $\sim 100$ kpc,
before coming to local pressure equilibrium.  At a speed of $356\rm\
km\ s^{-1}$ (see below), that would take $\sim 50$ Myr, leaving a
narrow window after it is disconnected from the jet in which cavity E
can continue to drive the shock.  A more plausible alternative is that
the cavities remain interconnected.  In that case, energy in the inner
cavities is only in transit and cannot be used to estimate jet power
(in the absence of a much more sophisticated model).  Only the
outermost cavities, which are the destination of the energy carried by
the jet, can be used to estimate its power.  If the northern jet remains
connected to cavity E, it is also easier to explain how the cavity
remains close to the shock front.

In these circumstances, the proximity of the shock to cavity E
suggests another estimate for the power of the northern jet.  
Radio losses are negligible, so that jet power is divided between
cavity thermal energy and $p\, dV$ work done by the expanding cavity, i.e.,
$P_{\rm jet} = dE_{\rm cav} / dt + p \, dV_{\rm cav} / dt$.
Assuming that evolution of the cavity and shock is approximately
self-similar, the radius of the cavity scales with the radius of the
shock, giving
$d R_{\rm cav} / dt \simeq v_{\rm shock} (R_{\rm cav} / R_{\rm shock})
\simeq 356 {\rm\ km\ s^{-1}}$,
for $R_{\rm cav} = 100$ kpc, $R_{\rm shock} = 350$ kpc, $kT = 3.2$
keV for the unshocked gas \citep{dnm01} and a shock Mach number of
1.34 \citep{nul05}.  Thus
$p dV_{\rm cav} / dt \simeq 4 \pi p R_{\rm cav}^3 v_{\rm shock} /
R_{\rm shock} \simeq 6.4 \times 10^{44} {\rm\ erg\ s^{-1}}$,
for the parameters above and a cavity pressure of $1.5\times10^{-11}
\rm\ erg\ cm^{-3}$.  If the pressure of the
cavity remains constant, including the increase in thermal energy
boosts this by $\gamma / (\gamma - 1)$, where $\gamma$ is the ratio of
specific heats for the lobe plasma.  Allowing for a
decrease in cavity pressure, $p \propto R_{\rm cav}^{-\eta}$, modifies
the boost to $\gamma / (\gamma - 1) - \eta/3$.  For self-similar
growth, the cavity pressure follows the external pressure and
$\eta\simeq 2$ from above, so that, for $\gamma=4/3$ in the cavity and
we get $P_{\rm jet} \simeq 2 \times 10^{45}\rm\ erg\ s^{-1}$.
Although evolution of the cavity and shock is unlikely to be exactly
self-similar, this estimate is better determined than those relying
on cavity rise times.

\section{History of the outburst}
\label{sec:history}

\newcommand\pcav{p}
\newcommand\vcav{V_{\rm cav}}

Jet power is divided between
thermal energy in the cavities, $\sim \pcav\vcav$,
and the work done by the cavities on the surrounding gas
as they expand, i.e., $\int \pcav \, d\vcav$ over the history of the
cavities (radiation losses are negligible).
The ratio $\pcav \vcav / \int \pcav\, d\vcav$ is then
a diagnostic for the history of an AGN outburst.
Increasing cavity volume and decreasing external pressure both tend to make
the cavity pressure decrease with time, so that jet power would need to
increase rapidly to maintain constant pressure.  Thus, unless the jet
power is increasing rapidly, $\pcav \vcav / \int \pcav \, d\vcav < 1$.
This ratio is smallest when the jet power is declining.

The observations do not give a clear indication of whether the
jets are steady, ramping up, or declining.
From \S~\ref{sec:energetics}, the total of $\pcav \vcav$ for the
cavities is $\sim 4\times10^{60}$ erg, which can be compared to the
total energy of $\sim 9\times10^{60}$ erg required for the shock model
\citep{nul05}.  In principle, the latter includes both the thermal
energy of the cavities and the work done by them.  This shock model is
spherical, is fitted to the part of the shock front closest to the
AGN, and assumes a
single, explosive energy injection event at the cluster center, rather
than continuous energy injection via jets.  These approximations tend
to underestimate the thermal energy of the cavities.  Nevertheless,
the numbers suggest
that the ratio $\pcav \vcav / \int \pcav\, d\vcav$ is not much smaller
than unity.  On this basis, the jet power does not appear to be  
declining rapidly at present.

On the other hand, the sharp bend in the northern jet at cavity C
indicates that the jet may not be momentum dominated at larger
radii\citep{nul05}.  If true, the long rise time of cavity E compared
to the age of the shock would require that it was partially inflated
in an earlier outburst.  The earlier outburst contributes to $\pcav \vcav$,
but not the current shock, so that $\pcav\vcav / \int \pcav\, d\vcav$
would be larger than expected for a single outburst.  Jet power may
then be declining more rapidly than suggested in the preceding
paragraph.  We note that the shape of the radio source and the
properties of cavity D are also consistent with a sharp bend in the
southern jet there (see \S~\ref{sec:energetics}).

\newcommand{\pjet}{P_{\rm j}}

There are other indications that the situation is more complex than
we might have imagined.
For example, cavities C and D may be kept
open by the forces required to deflect the jet.  If the total jet
power is $\sim2\times10^{45}\rm\ erg\ s^{-1}$ (\S~\ref{sec:ages}) and
half of this passes through the northern jet, then its minimum
momentum flux is $F = \pjet/c \simeq 3\times10^{34}\rm\ dyne$.  The
force required to deflect the jet through an angle $\theta$ is then $2
F \sin\theta/2 \simeq F$, for a deflection of $60^\circ$.  If the
outer wall of cavity C is cylindrical, with radius of curvature $r
\simeq 50$ kpc and depth $2r$ (cf. Table~\ref{tab:cavities}), the
minimum force per unit area on the cavity wall needed to deflect the
jet is $P_{\rm def} \simeq F/(\pi r^2/3) \simeq 1.3\times10^{-12}\rm\
erg\ cm^{-3}$.  This value is increased by a factor of $2c/v$, if the power
of the jet is primarily kinetic energy and its bulk speed is $v \ll
c$.  The gas pressure around cavity C is $\simeq 5\times10^{-11}\rm\
erg\ cm^{-3}$.  Thus, the force required to deflect the jet could keep
cavity C open if $v \simeq 0.1 c$.

Another issue for the history of the outburst is formation of the
innermost cavities.  In order to open them and/or keep them open, the
pressure in cavities A and B must equal or exceed the pressure of the
local ICM.  Furthermore, any recent pressure increase (change) must be
slow in order to avoid creating new shocks around these cavities.
Thus, if the innermost cavities were created by a recent boost in AGN
power, then it needs to have been gradual, over a timescale
significantly longer than their sound crossing time of $\sim30$ Myr
(Table~\ref{tab:cavities}).  The pressure in the ICM around cavities A
and B is also an order of magnitude greater than that around the
outermost cavities.  Since the pressure changes are gradual, we can
apply Bernoulli's theorem to the jet (whether relativistic or not) and
the large pressure change requires the bulk speed of the jet flow to
become highly supersonic (until a shock is encountered).

In summary, it is unclear whether the existence of multiple cavities
requires multiple outbursts.  The issues raised here highlight the
need for more detailed models in order to investigate the history of
this system.

\section{Discussion}
\label{sec:discuss}

A deep Chandra image of Hydra A has revealed a complex cavity 
system in the central 300 kpc of the Hydra A cluster indicating a long
and equally complex AGN outburst history energetically equivalent to
that of a powerful quasar.
The AGN has been active either continuously or intermittently
for the past several hundred million years, as it has deposited
several $10^{61}$ erg into the surrounding cluster gas.  
The corresponding jet power $2\times 10^{45}\rm\ erg\ s^{-1}$ is
enough energy to offset radiation losses associated with a several
hundred $\msunyr$ cooling flow \citep[eg.][]{dnm01}, and to heat the
gas well beyond the cooling region.  
The cavities occupy $\sim 10$ percent of the volume within 300 kpc of
the center, and they lie within about 40 degrees of the plane of the
sky. The cavities are filled entirely by 330 MHz radio emission,
demonstrating that low frequency radio emission faithfully traces 
the energy of the outburst.

The association between the observed radio emission and the X-ray
cavities implies that these cavities were created and powered by
accretion onto a supermassive black hole embedded in the cD.  
The inferred mass of the black hole based on the relationships between
black hole mass, stellar velocity dispersion, and cD galaxy luminosity
corresponds to $\sim 9\times 10^8 \msun$ \citep{rmn06}.  
The power, $P$, liberated over the lifetime of the event implies
an accretion rate of $P/\epsilon c^2 \sim  0.05-0.25 \msunyr$, where
we have adopted $\epsilon=0.1$ as the mass to energy conversion
efficiency.  The range of values depends on the adopted
age and total energy of the outburst.  
By comparison, the Eddington rate for a black hole of this mass is
$\simeq 20 \msunyr$, implying the accretion driving the outburst is in
the strongly sub-Eddington regime. 
The Bondi accretion rate implied by the black hole mass and central
gas density taken from \citep[][]{rmn06} is only $\sim 4 \times
10^{-4}\msunyr$, which is too small to power an outburst of this
magnitude. 
On the other hand, there is no shortage of cold fuel near the nucleus.
Approximately $ 10^4 - 10^7 \msun$ of neutral hydrogen is present in 
a circumnuclear disk or torus located within 5-30 pc of the nucleus
\citep{taylor96}.  
This material is enough to fuel the outburst for several tens of
thousands to several hundred million years.  Additional hydrogen and
CO is likely present in a larger circumnuclear disk of gas and star
formation \citep{mcn95} in amounts that do not exceed $\sim 10^9 \msun$
\citep{edge01}.

\acknowledgments

This research was supported by Chandra General Observer Program
grant GO4-5146A, and by NASA Long Term Space Astrophysics
grant NAG4-11025.

%
%

\end{document}